# Why is it so hard to find a job now? Enter Ghost Jobs[*]

Hunter Ng
Baruch College, City University of New York
hunterboonhian.ng@baruch.cuny.edu

29th October 2024

## Abstract

This study investigates the emerging phenomenon of "ghost hiring" or "ghost jobs", where employers advertise job openings without intending to fill them. Using a novel dataset from Glassdoor and employing a LLM-BERT technique, I find that up to 21% of job ads may be ghost jobs, and this is particularly prevalent in specialized industries and in larger firms. The trend could be due to the low marginal cost of posting additional job ads and to maintain a pipeline of talents. After adjusting for yearly trends, I find that ghost jobs can explain the recent disconnect in the Beveridge Curve in the past fifteen years. The results show that policy-makers should be aware of such a practice as it causes significant job fatigue and distorts market signals.

**Keywords**: Ghost jobs; Labor market dynamics; Job search fatigue; Glassdoor; ChatGPT; BERT; Beveridge Curve

**JEL Classifications**: J60, J23, J31, M51, C81, L22, D83, J45, J70, J11.

---

[*]I am grateful to Kevin Shih, Ruth Milkman, Lilia Maliar, Sandra Sequeira and an anonymous referee for helpful comments.

# 1. Introduction

Workers who apply for jobs assume that employers are looking to fill the jobs. That is not the case. Popular media has documented a new phenomenon of "ghost jobs"[1,2] (Paradis, 2024; Woods and Wong, 2024). Employers put out job offers but are actually not seeking anybody to fill the role. This is troubling for two reasons. Firstly, job-seekers incur costs to apply to positions and cause job-search fatigue (Lim et al., 2013), which can lead to longer-term unemployment. Secondly, employers creating a false market impression that a specific industry is thriving can actually divert economic resources wrongly (Goldfarb et al., 2007).

What is in it for the employers? I speculate three reasons. Firstly, human resource (HR) departments need to maintain a sense of work and continuity. If the company is fully filled, HR departments may be downsized as there is no significant work to do. Thus, to appear efficient, HR department engage in "productivity theater" (Sprague, 2023; Marchese, 2023). Secondly, constantly advertising vacancies is a strategy to combat long hiring times in specialized jobs (Woods and Wong, 2024). If HR knows that a job is difficult to fill, and if an employee suddenly quits and they cannot fill it in time, they may be forced to take the blame. Thirdly, receiving applications from the market constantly gives the firm an "on-the-ground" sense of the job applicants, their expected salaries and other advantageous information. They can use this information to keep current employees on edge if the market has an oversupply of candidates, or to get a sense of the current salary expectations for each specific skill set, and so on. This alludes to the opportunistic behavior of managers, which is well documented in the finance and accounting literature (Chalmers et al., 2002; Nienhaus, 2022).These result in the phenomenon of openings which have no intent of filling.

---

[1] In an interview done for the Connecticut Public Radio, the authors interviewed Lisa Simon, chief economist of Revelio Labs, a jobs analytics firm and she gave statistics that in 2018, 1 in 5 job postings would not result in a hire but in 2024, 1 in 2 job postings do not result in a hire. They cited an example Allison Giddens, owner of a company called WinTech, which makes parts for planes and the military. Giddens noted that she kept an ad all year round because it's hard to find the highly skilled workers she needs.

[2] A ghost job is distinct from ghost work, a term popularized by Gray and Suri (2019) which refers to hidden labor behind digital tasks and which if often exploited due to a lack of formal employment status.

In this article, I define an opening as a ghost job if it does not have an immediate or reasonably foreseeable immediate vacancy[3]. This definition includes jobs which the employer eventually wants to fill such as specialized jobs where the available labor pool is already small and firms want to put out ads to keep a pipeline of talent. Jobs which take a long time to hire, which may be due to stringent checks of job candidates do not fall under this definition.

To further illustrate this article's point, the trend of ghost jobs also explains discrepancies seen in Bureau of Labor Statistics (BLS) statistics. The Beveridge curve that finds a traditional relationship between job vacancies and other labor market indicators has become unstable in the last fifteen years (Mongey and Horwich, 2023)[4]. Figure 1 shows this increasing disconnect in the Beveridge curve in the last fifteen years. To further highlight this disconnect, the Job Opening and Labor Turnover Survey (JOLTS), which is a monthly report by the BLS on the US labor market, does not account for ghost job openings under its current definition[5]. Under its existing survey framework, the definition of an "opening" has three criteria. One of them states that the job "could" start within 30 days, which is unverifiable. Another states that the employer is "actively recruiting". Under this definition, a company that puts out a job ad, would have that opening qualify under this framework. Hence, from a policy standpoint, ghost jobs are counted in current vacancy rate surveys.

In this article, I study and characterize the existence of ghost jobs using a novel dataset of job interviews from Glassdoor. Glassdoor is a popular online platform that allows job-seekers to post their interview experiences. Statistics show that seventy percent of unemployed Americans use the Internet to search for jobs[6]. Glassdoor is a credible source of information that has been cited in previous empirical studies (Kim and Ra, 2022; Kyiu et al., 2023; Symitsi et al., 2018) and the ubiquity of the Internet in the job search process also lends statistical power to the Glassdoor data (Sockin and Sojourner, 2023).

---

[3]Ghost jobs have been defined on Wikipedia as a position that is non-existent or has already been filled. This definition does not conflict with the definition in this article but I add on to distinguish from the JOLTS definition

[4]Simon Mongey and Jeff Horwich provided a thought-provoking article in December, 2023 where they warned of a soft-landing using JOLTS data, contrary to what prominent economists such as Larry Summers and Olivier Blanchard have posited. They find that vacancy rates are not matching the other stock flow rates such as quitting rates and hiring rates, and they do not take a stand on the issue. However, they acknowledge that factors such as changes to'cost to employers of job posting, recruiting and evaluating candidates'

[5]A helpful anonymous referee pointed out that JOLTS's current survey framework does not capture ghost jobs.

This article relies on the *"Interviews"* section, which is displayed on each company's landing page in Glassdoor. A screenshot of the web interface is given in Appendix A1. I determine that a robust way to check whether a company is putting out a ghost job is to rely on word-of-mouth from past interviewees. While it is possible that some interviewees may simply be upset with a rejection from the company and write "sour grapes" comments about the company, the findings in this paper show that there is a distinction between such "sour grapes" comments versus evidence of ghost jobs. The "sour grapes" comments typically show displeasure with the process. For example, on 5th August 2022, a review posted for the position of *Sales Associate* at Robert Half stated -

*"An informational interview is followed by a video interview. I had multiple rounds of video interviews and I thought I did pretty well. As I am an immigrant, I guess they had some issues with our communication skills".*

Contrast this with a review posted on 29th April 2016 for the position of *Marketing Associate* at HDR -

*"It was a pretty slow process overall. I would say very slow. I mean they didnt call me back for 3 months. It's probably the worst HR ive ever encountered (I mean Omaha HR not hiring administrative staff). All the hiring is centered in Omaha. ... They didn't ask any cases, any "what would you do" questions, they didn't ask teamwork questions either. Pretty confusing overall. I initiated the salary negotiation myself because at that point I felt like they were going to hire me, but no one raised that topic."*

that may indicate that this was a ghost job, where there were no questions asked on competence, nor clarification on expected salary. More examples can be found in Appendix A1. There is a distinct difference in the content of "sour grapes" review versus "ghost job" reviews, which lends credibility to the dataset. Specifically, I analyze jobseekers' posts about their interview experiences through an advanced large-language model (LLM) textual analysis with AI-assisted techniques.

---

[6] Statistics taken from Sockin and Sojourner 2021's analysis of November 2019 Current Population Survey

My results find that a significant portion of up to 21% of job postings could indicate "ghost hiring". In line with my hypotheses, I also find that bigger firms with a bigger HR size have more ghost job ads although interesting, the large middle-sized firms tend to have the most ghost job ads; specialized jobs have more ghost job ads and lastly, . These results are robust to company and year fixed effects. By pooling data across multiple firms and industries, this article points to an uptick in trend of ghost jobs. Lastly, I also perform a time-series analysis using JOLTS data and reconcile with Mongey and Horwich's observation where vacancy rates are mismatched with unemployment data.

This article contributes to two streams of literature. Firstly, there is a stream of literature on classical labor search models and the disconnect of the Beveridge curve. Mortensen and Pissarides (1994) show a classic labor search model which has been the workhorse of unemployment modelling. Their model does not take into account job-seekers' cost due to fatigue from applying to ghost job openings. They also do not take into account information acquisition gains for firms putting out ghost jobs. Domash and Summers (2022) show that demand-side indicators, such as vacancy and quit rates, signal a tight U.S. labor market, with unemployment as a stronger predictor of wage inflation, indicating likely ongoing inflationary pressures. Mongey and Horwich (2023) posit that new technology advancements have affected the way vacancy rates traditionally show up in economic models such as the Beveridge curve.

Secondly, there is literature on LLM techniques being used to study economic phenomena. Jelveh et al., (2024) use machine learning to measure the partisanship in academic economics articles. Ziems et al., (2024) perform extensive testing which show that LLM's performance rivals humans in classifying and explaining social phenomena. Dell (2024) reviews how economics research stands to gain from deep neural networks which includes applications such as classifi-cation, document digitization, record linkage and methods for data exploration in massive scale text and image corpora.

This article adds to the literature by highlighting the practice of ghost jobs, which has not been extensively documented before. While the novel dataset used does not explicitly verify whether a job is a ghost job or not, a LLM-BERT is used, and the estimate is within range of an industry expert's statistics. It also uses first-hand accounts posted by verified job-seekers, something which was difficult to do empirically. It also offers an angle to reconciling the Beveridge curve disconnect highlighted by Mongey and Horwich. Ghost jobs also leads to other effects such as opportunistic companies posting ghost jobs to give a false impression of a booming performance, worsening job search fatigue for job-seekers, and misallocation of job search efforts and resources. The distorted labor market signals can also affect both job-seekers' expectations and the accuracy of labor market data used by policymakers and researchers. This is of concern to policy-makers and also to practitioners.

The remainder of this paper is structured as follows: Section 2 reviews relevant literature on ghost jobs and develops the hypotheses. Section 3 describes the methodology and results. Section 4 concludes with a summary of my key insights and suggestions for future work.

## 2. Background and Data

**2.1 Technology drives down the cost of posting job ads**

The idea of jobs that were not intended to be fill can be roughly traced to an article in 2013 when the Wall Street Journal first highlighted "phantom job listings" (Weber and Kwoh, 2013). In the article, Weber and Kwoh relied on anecdotes such as colleges posting ads as a "formality" and HR departments posting ads or risk losing "federal grant money". A survey of 1,045 hiring managers in 2022 showed that only 16% of jobs posted plan to be filled in one month (Clarify Capital, 2023). The prevalence of the trend shows a loophole in JOLTS, which cannot verify the authenticity of these jobs but nonetheless, classifies them as legitimate job openings. This has led to a widening disconnect and irrelevance in the Beveridge curve, where unemployment rate is inversely proportional to job vacancies.

A salient reason driving the increase in ghost jobs is the minimal marginal cost to employers in posting a new job on top of existing job postings (Bhuller et al. 2023)[7][8]. This is different from the hiring costs in classical labor economics literature, which entails other costs such as interview costs, writing the job description, processing the salary and benefits, on-boarding costs, etc. Table 1 shows a comparison among the top ten popular job portals and that there is next to no marginal cost to posting a new job ad.

While the phenomenon of ghost jobs has been showcased through anecdotes and surveys, empirical evidence is lacking due to the difficulty in verifying whether an opening is a ghost job or not. While one could get data on how many jobs are posted and left for a period of more than 30 days or are continuously posted throughout the year, these cannot robustly be categorized as ghost jobs because the hiring firm can argue that they have difficulty finding good candidates. In the next section, I introduce the novel dataset and an AI-assisted method to classify the data.

## 2.2 A LLM-BERT approach to classifying reviews as indicative of ghost jobs

Glassdoor is a popular online platform founded in 2007 to allow employees to anonymously share reviews about employers. Over time, it has become a major resource for job seekers and researchers. In this article, I focus on the pre-employment interview section that allows job-seekers to post their interview experiences. A job seeker can submit one interview review per employer per year on Glassdoor, which they can update as needed. However, they can review the interviews of multiple employers within the same year. Each review pertains to a specific job at a particular firm. Each review has an associated volunteer (the reviewer) and a creation time.

---

[7] In the same interview in Footnote 1, Giddens noted that job listing companies gave her an incentive to put up more job listings than are immediately required. She cited ZipRecruiter, a recruiting company charging her $400 a month for three job listings, even though she only needed one hire. The rational mindset was thus - since the job ad was already paid for, there was no marginal cost to posting two more, even though she was not looking to fill them.$

[8] Appcast, which is a subsidiary of Stepstone Group, the 3rd largest jobs business in the world behind Indeed and LinkedIn, documented in its annual white paper that cost per click of hiring has been consistently decreasing since 2021.

Figure 4A shows the survey users fill out when submitting an employer interview review to Glassdoor. Fields marked with an asterisk are required. The overall rating is restricted to whether it was positive, neutral or neutral. They then add the job title, interview difficulty, process and other interview questions.

To create accountability, Glassdoor requires reviewers to have a verified active email address or a valid social network account. Each submitted review is assessed for compliance with content guidelines, and those that do not meet the standards are not displayed. Every interview review posted is anonymous but their accounts are verified. The review also shows whether the job-seeker was employed or not, and there is also an option to indicate the location of the interview. Figure 4B shows such an example.

Once an interview review is posted, visitors to the website can see it, vote it as helpful, and answer or comment about the interview questions posted. If they find that the interview review is problematic, they can also report it to Glassdoor, which will trigger an internal investigation.

When a job-seeker reads a review, they can click and label whether the review was helpful or refrain entirely from expressing an opinion. Glassdoor also provides additional information about reviewed firms through its list of companies page. I collect this information and the list of variables is shown in Table 1.

To build my dataset, I extracted the first 1,203 companies listed by relevance from Glassdoor. As shown in Appendix A1, I use the filter where overall rating is 1.0 and above to capture as many companies as possible. I then extract 30 pages of interview reviews from each company's Glassdoor page. The web-scraping is done in the month of July 2024 and takes a month to complete. I retrieve a total of 29,294 pages from 1,203 companies. Some companies are smaller or not as popular and may not have up to 30 pages of interview reviews. There are also some reviews not in English as Glassdoor is an exhaustive dataset of global companies and MNCs. I use a deep-learning, language detector library in Python and identify that there are 269,347 english reviews, 4,315 portugese reviews, 3,593 french reviews and etc. The full list is given in Appendix A2. Table 2 shows the descriptive statistics of the variables. I use only the english reviews and have 1,199 valid companies across 97 industries.

To identify job interview reviews as indicative of ghost jobs, I first use a natural language processing ("NLP") technique of Latent Dirichlet Allocation ("LDA"). Preliminary results show that the interview reviews are high-quality and reflect logical issues with the interview process.

Next, to label whether an interview review is indicative of a ghost job or not, I use two different approaches. In the first approach, I manually conceive a list of key words that could be indicative of ghost jobs. I then use a bag-of-words approach, where if the interview experience contains these words subject to prior data cleaning, I label these as indicative of "ghost jobs". This approach follows classical NLP techniques used in the literature (Loughran and McDonald, 2011).

In the second approach, I use an AI-assisted approach to classify the interview experiences. Using the model of ChatGPT-4o, I upload the reviews to perform the classification. Using this output, I train a BERT model on this data to then classify whether the job review is indicative of a ghost job or not. I use this mixed approach because the dataset of 269,347 would be too costly to classify entirely with ChatGPT-4o, while the mixed approach offers a faster and cost-effective solution.

As the two approaches differ in how they determine whether a review is indicative of ghost jobs, I hypothesize that the BERT model would be more robust. This is because a plain search of key words may not provide enough context of ghost jobs as such textual clues could be embedded within the review itself. Thus, I state my first hypothesis in alternate form.

**H1: A BERT model would better identify reviews indicative of ghost jobs than a keyword search**

**2.3 The Beveridge Curve disconnect**

One of the economic issues in this article revolve around the Beveridge Curve. It is a concept used to explain the relationship between job vacancies and unemployment. Over time, the curve has become disconnected where high vacancies persist alongside higher unemployment. Mongey and Horwich (2023) take a neutral position that technological trends may have influenced this disconnect. I posit that one of the contributing factors could be the rise of ghost jobs. Referring to Figure 1, the disconnect has been growing every year. This can complicate the analysis of the labor market and economic policy.

**2.4 Highly-skilled jobs are more likely to be ghost jobs**

Next, based on the literature, highly-skilled jobs are more likely to be ghost jobs because they draw from a smaller labor pool and take longer to fill. Unlike lowly-skilled positions that have a larger number of potential applicants, highly-skilled roles often require specialized expertise, higher education qualifications, and extensive experience. This limited availability of qualified candidates means that employers face more significant challenges in finding suitable hires quickly. As a result, firms keep job advertisements open for extended periods to ensure they capture the interest of highly-skilled professionals whenever they become available. This allows companies to remain competitive by having a steady stream of potential candidates for critical positions, thereby reducing the time it takes to fill these roles when vacancies arise.

Moreover, the longer time required to replace high-skilled jobs further incentivizes employers to maintain a continuous pipeline of applicants. In industries where highly-skilled positions are vital, sudden vacancies can severely disrupt operations and project timelines. By advertising these roles even when there is no immediate need, firms can conduct preliminary screenings and build a database of potential hires. This approach minimizes the downtime associated with the hiring process, especially for jobs in certain environments (Moretti, 2013). Thus, I propose my hypothesis in the alternative form:

**H3: Highly-skilled industries are more likely to have ghost job postings**

**2.5 Bigger firms are more likely to have ghost jobs**

Bigger firms are also more likely to have ghost jobs due to their extensive resources and complex organizational structures. Large companies often have multiple departments and a constant need to fill a variety of positions, making it more feasible for them to keep job advertisements open even when there are no immediate vacancies. This maintains a robust talent pipeline, ensuring that they can quickly respond to unexpected departures or sudden increases in demand. Additionally, bigger firms typically have dedicated HR departments with the capacity to manage ongoing recruitment processes, including screening and maintaining databases of potential candidates. There is also literature on how operational factors can affect stock returns, and the high vacancy rate may positively signal to shareholders that the company is thriving (Campbell, 1996; Chen and Li, 2023).

Moreover, larger organizations benefit from economies of scale in their hiring processes, making the marginal cost of maintaining job ads relatively low. The job portals often offer bulk advertising deals or subscription-based models that allow companies to post numerous job listings at a reduced cost. This low marginal advertising cost makes it practical for big firms to advertise positions continuously, even if they are not actively looking to hire immediately. The presence of ghost jobs in larger firms can also be attributed to strategic workforce planning, where companies aim to gather data on the labor market, including salary trends and skill availability. By keeping job ads active, they can continuously monitor market conditions. Thus, I hypothesize the following in alternate form.

**H4: Bigger firms are more likely to have ghost job postings**

## 3. Results

**3.1 Addressing the Beveridge Curve Disconnect - What is the percentage of job ads that are ghost jobs? What are the characteristics and sectors of these ghost jobs?**

To begin, I extract the review data from Glassdoor. I then remove the non-english reviews from my initial sample. I remove the carriage returns and perform a LDA topic analysis. To do so, I first tokenize the reviews using the bag-of-words approach and then use the *gensim* package to build the LDA model. Each LDA topic is represented as a mixture of words, and each word in the corpus is assigned to one of the topics. The results are shown in Table 3 Panel (A). The results confirm that the Glassdoor data is high-quality. From the 10-word, 5 topic LDA, the interview reviews legitimately focus on issues common to interviews, such as the process with HR, hiring timeline, question difficulty, etc.

Next, to classify whether the reviews were indicative of ghost jobs or not, I first used a keyword search model. I hand-compiled the list of keywords from the news articles documenting the ghost jobs trends and also through manual examination of the reviews.

In the second AI-assisted LLM approach, I use the same 2000 interview reviews and used ChatGPT-4o, an advanced artificial intelligence model to classify whether they were indicative of ghost jobs or not. The prompts and methodology are listed in Appendix A2. After the classification, I then trained a BERT model to the entire dataset. The specifications of model-training in Appendix A2.

From the results in Table 3 Panels (B) and (D), I see that there is a huge difference if I use a deep-learning BERT model to distinguish whether an interview review indicates ghost jobs. Through comparison of the results where 1.6% and 21% of the total reviews are ghost jobs for the keyword-search and BERT respectively , I conclude that the LLM-BERT model performs better at classifying whether a review is indicative of a ghost job. The 21% cited is much closer to the expert statistic of 50% by Lisa Simon (Woods and Wong, 2024). I acknowledge that while some of these may simply be that the company has hired another better candidate, thus biasing my results upwards, there are also many other job candidates who may not have written a review about the ghost job trend, which bias the results downwards. Although the keyword search is a classic method of locating themes in the review, it may be lacking because there is a lot of peripheral meaning that needs to be inferred when concluding whether an interview review is indicative of a ghost job. I further give some examples in Appendix A1 that indicate ghost jobs but would not show up in the keyword search.

To reconcile the Beveridge Curve disconnect, I plot a line graph of the number of indicative ghost jobs for the dataset. Figure 2 shows the plot, and there is a jump in 2015. This can be attributed to Glassdoor receiving 70 million USD in funding, which increased its reach and US advertising (Cook, 2015). Barring this jump, Figure 2 also follows the trend in Figure 1 during 2020, where Covid caused the fall in both *job quit rate* and *job openings rate*. I take this to show that the prevalence of ghost jobs highlighted through Glassdoor interview data became more reflective of the actual trend following this expansion in 2015.

To integrate the ghost jobs dataset to JOLTS's data, I use a rolling average adjustment to smooth the addition of indicative ghost jobs on JOLTS *job openings rate*, reducing the influence of short-term fluctuations while retaining sensitivity to longer-term trends. A rolling average of indicative ghost jobs over each year stabilizes the data. To align the adjustment with the typical gap between JOLTS *job openings rate* and JOLTS *job quit rate*, I apply a dynamic scaling factor based on this average gap. This makes for a smoother and more consistent adjustment, helping maintain a stable relationship between the openings and quits rates over time.

Figure 3 shows the adjusted openings rate. We can see that with the exception of the Covid period, which could have affected the posting of interview reviews, the Beveridge curve starts to close back up after 2022. This shows potentially that ghost jobs could offer an explanation to the discrepancy between jobs opening and jobs quit rate, restoring the validity of the Beveridge Curve.

I then continue to perform a LDA on the interview reviews that have been classified as ghost jobs by both models. I present the results in Table 4. From Table 4 Panel (A), we see that the top five topics revolve around the experience of the interviewee, where there are multiple rounds of interview, long waiting times, lack of salary negotiations and no follow-up. Table 4 Panel (B) continues to show that negative experiences stand for the highest percentage of indicative ghost jobs, which show that the BERT model correctly classifies the experiences. 74.8% of negative experiences are indicative of a ghost job under the BERT model, while 6.23% of the negative experiences are indicative of a ghost job under a keyword search.

**3.2 Do small or big firms push out more ghost job ads? Do higher skilled industries push out more ghost jobs?**

Next, I want to understand the cross-sectional characteristics of ghost jobs. From the evidence, I hypothesize that the bigger a firm, the more likely there is slack in the HR or that there is a bigger need to gain more information from the market constantly about expected pay, skill-set or other useful data. From the sample, I split the companies into 7 distinct categories ranging from 1 to 50 employees all the way to 10,000 employees. I then calculate the number of ghost jobs and the percentage of ghost jobs based on the number of interview reviews indicative of ghost jobs divided by total number of interview reviews of companies in that employee number bracket. Table 5 presents these results.

Next, I hypothesize that industries with more specialized and difficult skills would have more indications of ghost jobs. I separated the interview reviews based on industry. There are 97 industries for the entire dataset. After grouping by industry, I calculated the highest and lowest % of reviews that were indicative of ghost jobs and present them in Table 6.

From Table 5, I find that the largest % of ghost jobs center around companies with a high number of employees. Specifically, the keyword search model reveals that 1.58% is the highest for companies with 5,001 to 10,000 employees and the BERT model is the highest for companies with 1,001 to 5,000 employees. This is consistent with our hypothesis that ghost jobs are most often put out by companies more employees, which may have excess capacity in their HR departments and job ad credits. The small and biggest firms are able to keep to leaner practices and spread out their resources respectively, leading to much lower %s of putting out ghost jobs.

Table 6 continues to confirm the hypothesis that high-skilled jobs are more likely to see trends of ghost jobs. Sectors such as publishing, Internet and web services, software development, and commercial equipment services all require highly skilled labor but also have a relatively large available pool of workers. These are also jobs that have been experiencing major hiring trends. For the charitable foundations, the fact that they often rely on government grants also strengthen the results.

Routine and less-skilled jobs such as restaurants have the lowest percentage of ghost jobs. Interestingly, national agencies, religious institutions also show a low percentage of ghost jobs. This could point to the fact that national agencies have no need to put up ghost jobs since there are no advantages to receiving federal grants as suggested by the Clarify Capital survey. Accounting and tax firms have a large labor pool to hire from so they do not engage in such practices. Religious institutions also see a low ghost jobs rate, probably because job-seekers in this area are already aligned with the faith and may not want to interpret their experiences as ghost jobs. The commercial printing sector has a relatively stable demand with predictable customer bases so companies are also probably less likely to engage in ghost jobs practice.

Publishing and software related industries consistently see high percentages of ghosting practices, possibly due to the fact that labor may be scarcer since they are niche areas and that there may be a more dire need to build a pipeline of talents.

**3.3 Does the professional or management nature of the job affect the likelihood of it being a ghost job?**

To investigate the nature of these ghost jobs, I first identify the word count of each interview review. Next, I use a keyword search to determine whether a job ad is catering to a highly skilled/ professional/management or lowly skilled/labor-intensive/trainee job. I then use a fixed-effect regression of the specification below to determine whether the variables are of significance or not. Table 7 shows the results.

Table 7 shows that even when in taking year and fixed effects, the more highly-skilled a job is, the more likely it is a ghost job. This offers strong evidence in line with my hypothesis. Additionally, column (1) shows significance in *size* and $size^2$, which corroborates Table 5 where there is a N-shaped curve in that the number of ghost jobs peak when firms reach middle size. In Appendix A3, I set up a model to show that using classical costly information acquisition assumptions, middle-sized firms are the ones possibly posting the most ghost jobs under certain conditions.

Skill also has a large effect on whether the job is indicative of a ghost job or not in Columns (1), (2), (3). This corroborates the results in Table 6 and shows strong evidence both quantitatively and qualitatively that highly-skilled jobs show more evidence of ghost-jobs. This result is also stronger given that there are naturally more lowly-skilled jobs then highly-skilled jobs (BLS, 2024), which bias the results in the opposite manner.

## 4. Conclusion

The phenomenon of ghost hiring, where employers advertise jobs without the intention of filling them, has significant implications for both job seekers and the broader labor market. This article, leveraging data from Glassdoor and advanced natural language processing techniques, provides robust evidence that ghost jobs are prevalent and can distort labor market dynamics.

Firstly, I find empirically that a considerable proportion of job ads are ghost jobs, with the LLM-BERT model estimating that 21% of the reviews indicate such practices. This prevalence is particularly high in industries requiring specialized skills and among larger firms, suggesting that these companies use ghost ads to maintain a continuous flow of applicant information and to manage employee expectations. The minimal marginal cost of posting job ads further worsens this trend, as employers face little financial deterrent to continually advertise vacancies. While there may be many reasons why the middle-sized firms post the most ghost jobs, I offer a theoretical model from an information acquisition point of view that middle-sized firms may be the ones that post the most ghost job ads.

Secondly, the impact on job seekers is significant as ghost jobs contribute to job search fatigue, increased application costs, and potentially longer periods of unemployment. These effects can skew job seekers' expectations and lead to a misallocation of resources, as they invest time and effort in pursuing positions that are not genuinely available. This, in turn, can elevate reservation wages and disrupt the equilibrium in job search models, affecting overall labor market efficiency.

Furthermore, my findings have broader implications for economic policy and labor market research. The distortion caused by ghost jobs can lead to inaccurate labor market indicators, affecting empirical studies and policy decisions based on these metrics. Policymakers need to consider the impact of ghost hiring practices when designing interventions aimed at improving labor market efficiency and reducing unemployment.

In conclusion, while employers might have strategic reasons for maintaining ghost jobs, the practice poses substantial costs to job seekers and can distort labor market signals. To solve this issue, perhaps greater transparency in job advertising practices and regulatory measures to ensure that job ads reflect genuine hiring intentions could help. Future research can continue to explore trend, including its long-term effects on labor market dynamics.

# References

Appcast. (2024). 2024 Recruitment Marketing Benchmark Report (US Edition). https://whitepaperseries.com/wp-content/uploads/2024/03/135886_ML-2024-Recruitment-Marketing-Benchmark-Report.pdf#page=1.00

Bhuller, M., Ferraro, D., Kostøl, A. R., & Vigtel, T. C. (2023). The Internet, Search Frictions and Aggregate Unemployment (Working Paper 30911). National Bureau of Economic Research. https://doi.org/10.3386/w30911

BLS. (2024). BLS: U.S. Bureau of Labor Statistics. Bureau of Labor Statistics. https://www.bls.gov/oes/tables.htm

Campbell, J. Y. (1996). Understanding Risk and Return. Journal of Political Economy, 104(2), 298–345. https://doi.org/10.1086/262026

Chen, C.-W., & Li, L. Y. (2023). Is hiring fast a good sign? The informativeness of job vacancy duration for future firm profitability. Review of Accounting Studies, 28(3), 1316–1353. https://doi.org/10.1007/s11142-023-09797-2

Chalmers, J. M. R., Dann, L. Y., & Harford, J. (2002). Managerial Opportunism? Evidence from Directors' and Officers' Insurance Purchases. The Journal of Finance, 57(2), 609–636. https://doi.org/10.1111/1540-6261.00436

Clarify Capital. (2023, October 23). Survey: Job Seekers Beware of Ghost Jobs. Clarify Capital. https://clarifycapital.com/job-seekers-beware-of-ghost-jobs-survey

Cook, J. (2015, January 6). Rich Barton-backed career site Glassdoor picks up 70 million in funding. GeekWire. https://www.geekwire.com/2015/rich-barton-backed-career-site-glassdoor-picks-70-million-funding/

Dell, M. (2024). Deep Learning for Economists (Working Paper 32768). National Bureau of Economic Research. https://doi.org/10.3386/w32768

Domash, A., & Summers, L. H. (2022). How Tight are U.S. Labor Markets? (Working Paper 29739). National Bureau of Economic Research. https://doi.org/10.3386/w29739

Goldfarb, B., Kirsch, D., & Miller, D. A. (2007). Was there too little entry during the Dot Com Era? Journal of Financial Economics, 86(1), 100–144. https://doi.org/10.1016/j.jfineco.2006.03.009

Gray, M. L., & Suri, S. (2019). Ghost work: How to stop Silicon Valley from building a new global underclass. Eamon Dolan Books. https://books.google.com/books?hl=en&lr=&id=8AmXDwAAQBAJ&oi=fnd&pg=PP1&dq=Ghost+Work:+How+to+Stop+Silicon+Valley+from+Building+a+New+Global+Underclass&ots=WVI6PT2Z2r&sig=bWHBrOQd1bMnXNf2Sbki9OVjKYk

Grossman, S. J., & Stiglitz, J. E. (1980). On the impossibility of informationally efficient markets. The American Economic Review, 70(3), 393–408.

Kim, J. (Simon), & Ra, K. (2022). Employee satisfaction and asymmetric cost behavior: Evidence from Glassdoor. Economics Letters, 219, 110829. https://doi.org/10.1016/j.econlet.2022.110829

Kyiu, A., Tawiah, B., & Boamah, E. O. (2023). Employees' reviews and stock price informativeness. Economics Letters, 233, 111406. https://doi.org/10.1016/j.econlet.2023.111406

Lim, V. K., Chen, D. J. Q., & Tan, M. (2013). Unemployed and Exhausted? Fatigue During Job-Search and Its Impact on Reemployment Quality. Academy of Management Proceedings, 2013(1), 13399. https://doi.org/10.5465/ambpp.2013.13399abstract

Loughran, T., & Mcdonald, B. (2011). When Is a Liability Not a Liability? Textual Analysis, Dictionaries, and 10-Ks. The Journal of Finance, 66(1), 35–65. https://doi.org/10.1111/j.1540-6261.2010.01625.x

Marchese, D. (2023, January 23). The Digital Workplace Is Designed to Bring You Down. The New York Times. https://www.nytimes.com/interactive/2023/01/23/magazine/cal-newport-interview.html

Mongey, S., & Horwich, J. (2023). Are job vacancies still as plentiful as they appear? Implications for the "soft landing" | Federal Reserve Bank of Minneapolis. Federal Reserve Bank of Minneapolis. https://www.minneapolisfed.org/article/2023/are-job-vacancies-still-as-plentiful-as-they-appear-implications-for-the-soft-landing

Moretti, E. (2013). Real Wage Inequality. American Economic Journal: Applied Economics, 5(1), 65–103. https://doi.org/10.1257/app.5.1.65

Mortensen, D. T., & Pissarides, C. A. (1994). Job creation and job destruction in the theory of unemployment. The Review of Economic Studies, 61(3), 397–415.

Morris, S., & Shin, H. S. (2002). Social value of public information. American Economic Review, 92(5), 1521–1534.

Nienhaus, M. (2022). Executive equity incentives and opportunistic manager behavior: New evidence from a quasi-natural experiment. Review of Accounting Studies, 27(4), 1276–1318. https://doi.org/10.1007/s11142-021-09633-5

Paradis, T. (2024). It's wild how many job listings might be fake. Business Insider. https://www.businessinsider.com/companies-posting-fake-job-listings-search-resume-2024-6

Sims, C. A. (2003). Implications of rational inattention. Journal of Monetary Economics, 50(3), 665–690.

Sockin, J., & Sojourner, A. (2023). What's the Inside Scoop? Challenges in the Supply and Demand for Information on Employers. Journal of Labor Economics, 41(4), 1041–1079. https://doi.org/10.1086/721701

Sprague, R. (2023). Privacy Self-Management: A Strategy to Protect Worker Privacy from Excessive Employer Surveillance in Light of Scant Legal Protections. American Business Law Journal, 60(4), 793–836.

Symitsi, E., Stamolampros, P., & Daskalakis, G. (2018). Employees' online reviews and equity prices. Economics Letters, 162, 53–55. https://doi.org/10.1016/j.econlet.2017.10.027

Vaswani, A., Shazeer, N., Parmar, N., Uszkoreit, J., Jones, L., Gomez, A. N., Kaiser, Ł., & Polosukhin, I. (2017). Attention is all you need. Advances in Neural Information Processing Systems, 30. https://proceedings.neurips.cc/paper/2017/hash/3f5ee243547dee91fbd053c1c4a845aa-Abstract.html

Veldkamp, L. (2023). Information choice in macroeconomics and finance. Princeton University Press. https://books.google.com/books?hl=en&lr=&id=qlSIEAAAQBAJ&oi=fnd&pg=PP9&dq=Information+Choice+in+Macroeconomics+and+Finance&ots=DNuPDbVFEL&sig=tA8ia0BC7g1DB5pVABPnD-auDBs

Weber, L., & Kwoh, L. (2016, May 12). Want a New Job? Beware "Phantom" Postings - WSJ. Archive.Ph. https://archive.ph/kNYmI

Woods, D., & Wong, W. (2024). What are "ghost jobs"? Connecticut Public Radio. https://www.ctpublic.org/2024-06-14/what-are-ghost-jobs

Ziems, C., Held, W., Shaikh, O., Chen, J., Zhang, Z., & Yang, D. (2024). Can Large Language Models Transform Computational Social Science? Computational Linguistics, 50(1), 237–291. https://doi.org/10.1162/coli_a_00502

# Tables and Figures

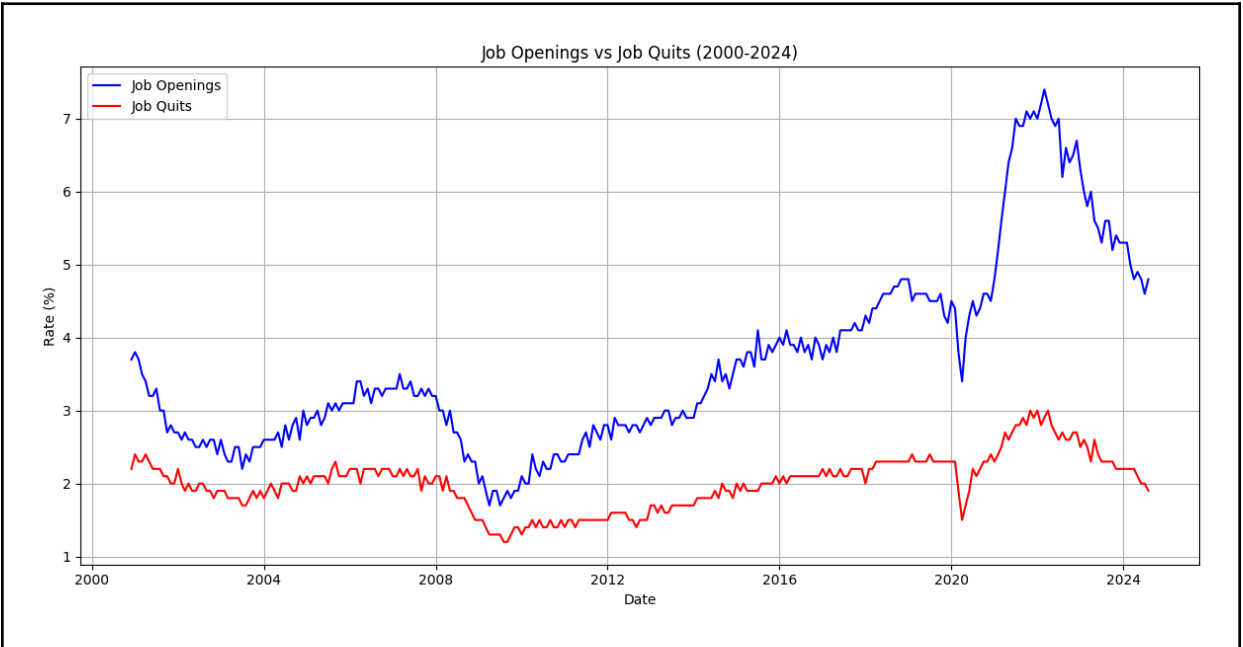

**Figure 1.** Comparison of seasonally adjusted *job openingrate* and *job quits rate* in the US. Source: Bureau of Labor Statistics, Jobs Openings and Labor Turnover Survey (JOLTS)

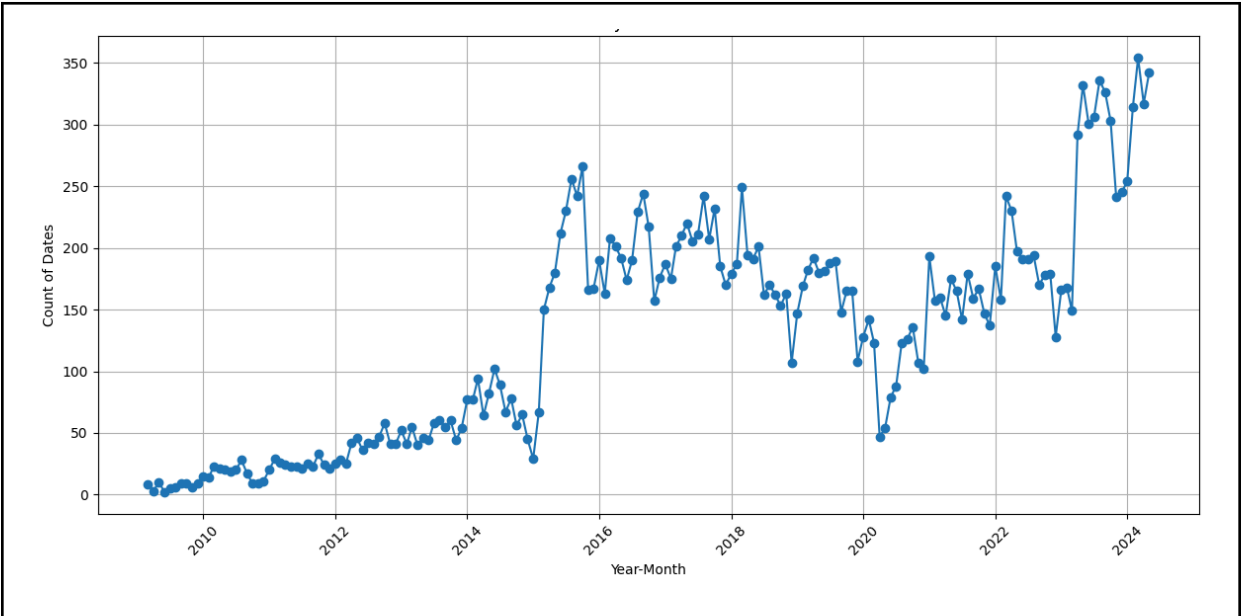

**Figure 2.** Number of ghost jobs classified by BERT for each month. Only interview reviews with a US location are taken.

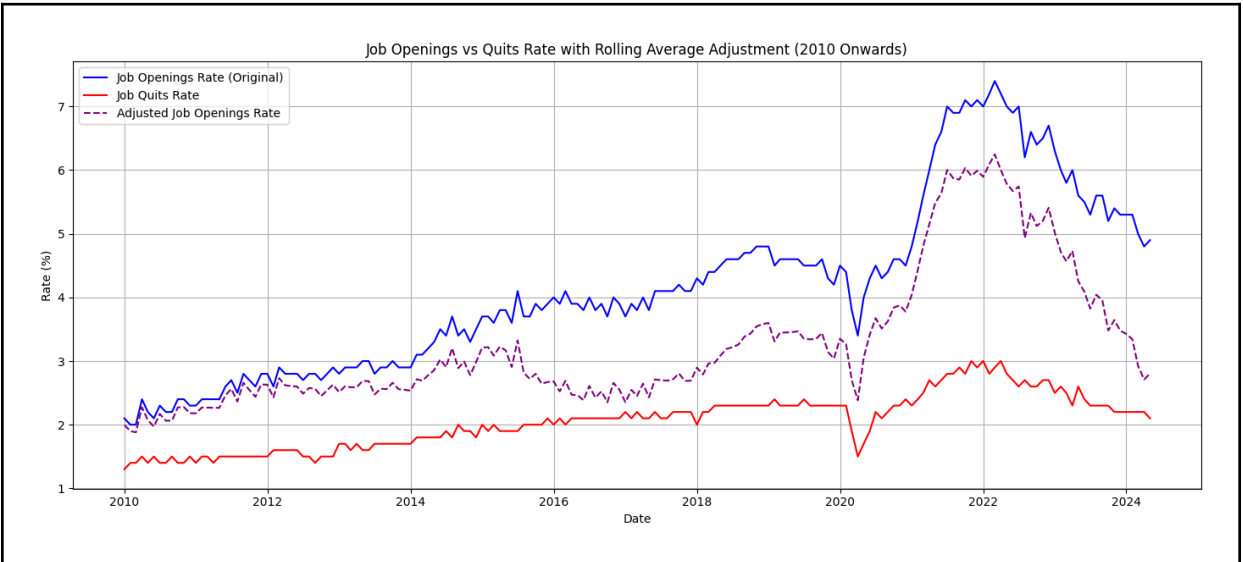

**Figure 3.** Rolling average-adjusted *job opening rate* after taking into account the number of ghost job reviews posted, and the *job quits rate* in the US. This line graph shows that the disconnect closes back up after the Covid pandemic.

**Table 1.** How much does it cost to post a job ad in 2024? Table 1 describes the cost of posting a job ad on the most popular online recruitment websites. There are many free options, which also feature job postings by SnP500 companies. This provides evidence that job ad posting costs has become marginal today. Source - various websites

| Jobs Portal | Cost | Details |
| --- | --- | --- |
| Indeed | 0 | Indeed doesn't charge for posting ads, only for showcasing your post |
| LinkedIn | 0 | First job posting is free, subsequent jobs charged based on cost-per-click |
| Glassdoor | 0 | Glassdoor lets companies post up to 10 jobs free for 7 days |
| Handshake | 0 | Unlimited posting for free |
| SimplyHired | 0 | Unlimited posting for free |
| Guru | 0 | Unlimited posting for free |
| HubStaffTalent | 0 | Unlimited posting for free |
| Illfound | 0 | Unlimited posting for free |
| ZipRecruiter | $16/day | Charged for jobs posting per day |
| Monster Jobs | $12/day | Charged for jobs posting per day |

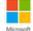

**Fig 4A.** Screenshot of the Glassdoor InterviewReview Section, where job-seekers can post their experiences.

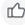

**Fig 4B.** Screenshot of an interview review posted for Louis Vuitton in New York on 3th April 2024

**Table 1.** Variable Definitions. This table presents definitions of the variables used in the paper.

| Variables | Definition | Source |
|---|---|---|
| $review_{i,t}$ | Interview Review, based on company $i$ and time $t$ | glassdoor.com |
| $experience_{i,t}$ | Sentiment of the Interview Review, whether it is good, neutral of negative, based on company $i$ and time $t$ | glassdoor.com |
| $word\_count_{i,t}$ | Length of words of Interview Review, based on company $i$ and time $t$ | glassdoor.com |
| $industry_{i,t}$ | Industry of the company revieId for the job interview based on company $i$ and time $t$ | glassdoor.com |
| $skill_{i,t}$ | Whether the job is a highly skilled, professional or managerial role. Takes a boolean of 1 for highly skilled and 0 otherwise based on company $i$ and time $t$ | glassdoor.com |

**Table 2**. Descriptive Statistics of the final sample size used in this study

|  | count | mean | min | 25% | 50% | 75% | max | std |
|---|---|---|---|---|---|---|---|---|
| *review* | 269,347 | 134,674 | | | | | | |
| *word_count* | 269,347 | 56.6 | 1 | 33 | 40 | 56 | 967 | 53.4 |
| *date* | 269,347 | | 3/11/2009 | 9/14/2017 | 6/25/2021 | 5/17/2023 | 7/16/2024 | |
| *skill* | 199,199 | 0.642 | 0 | 0 | 1 | 1 | 1 | 0.479 |
| *company* | 1,199 | | | | | | | |
| *industry* | 97 | | | | | | | |

**Table 3 Panel A.** Latent Dirichlet Analysis on the interview reviews. Sample size = 269,347.

| Topic | Keywords | Interpretation |
| --- | --- | --- |
| 0 | 0.019*time, 0.014*would, 0.013*recruiter, 0.012*company, 0.011*candidate, 0.010*even, 0.009*interview, 0.009*process, 0.008*role, 0.008*like | Focuses on the candidate's perspective, including interactions with recruiters and company roles in the interview process. |
| 1 | 0.069*round, 0.046*interview, 0.045*question, 0.037*technical, 0.020*hr, 0.013*asked, 0.013*first, 0.013*one, 0.012*2, 0.011*coding | Focuses on the different rounds of interviews, particularly technical and HR interviews, and specific questions such as coding |
| 2 | 0.034*interview, 0.030*question, 0.021*test, 0.018*online, 0.018*assessment, 0.015*take, 0.015*answer, 0.013*time, 0.012*video, 0.011*minute | Focuses on the structure and components of the interview process, including online assessments and video interviews. |
| 3 | 0.111*interview, 0.038*manager, 0.027*phone, 0.027*call, 0.024*process, 0.022*hiring, 0.021*recruiter, 0.020*week, 0.016*team, 0.015*day | Focuses on the managers and recruiters, phone interviews, and the overall hiring timeline. |
| 4 | 0.050*interview, 0.044*question, 0.025*process, 0.024*experience, 0.019*asked, 0.017*good, 0.016*easy, 0.013*job, 0.013*work, 0.012*interviewer | Focuses on the overall experience of the interview process, question difficulty and interviewer interaction. |

**Table 3 Panel B.** Keyword Matching to check whether interview review indicative of ghost jobs.

| Category | Count |
| --- | --- |
| Number of reviews matched for any of the keyword in the list | 3,624 |
| Number of reviews not matched for any of the keyword in the list | 265,723 |
| % of Ghost Jobs | 1.3% |

**Table 3 Panel C.** List of keywords matched to the interview reviews and their count. Note that this algorithm only matches the first word that is matched to the list of keywords. For example, a sentence with "ghosted" and "no intent" would be subsumed under "ghosted" as it is first in the list.

| Keyword | Count |
| --- | --- |
| ghosted | 3,169 |
| position had been filled | 125 |
| no intent | 68 |
| no longer hiring | 34 |
| just interviewing | 29 |
| no job offer | 25 |
| no intention of hiring | 22 |
| position was put on hold | 19 |
| Others | 133 |
| Total | 3,624 |

**Table 3 Panel D.** BERT Matching to check whether interview review indicative of ghost jobs

| Category | Count |
| --- | --- |
| Number of True values in 'keyword-match' column for English reviews | 56,559 |
| Number of False values in 'keyword-match' column for English reviews | 212,788 |
| % of Ghost Jobs | 21% |

**Table 4 Panel A.** Latent Dirichlet Analysis on the interview reviews which have been classified as ghost jobs using the BERT model. Sample size = 56,559.

| Topic | Keywords | Interpretation |
|---|---|---|
| 1 | interview, round, process, hr, time, 2, feedback, 3, final, month | Focuses on multiple rounds, HR involvement, feedback, and the time frame of the interview stages. |
| 2 | interview, job, day, manager, time, one, told, get, minute, would | Focuses on individual interview experiences, including interactions with managers, the time taken, and what candidates were told during the process. |
| 3 | interview, recruiter, call, email, week, back, would, never, time, phone | Focuses on interactions with recruiters, follow-up calls and emails, and instances of delayed or no responses. |
| 4 | company, like, time, work, know, candidate, even, get, people, want | Focuses on perceptions and experiences with the company, including the work environment and how they felt about the company's communication and treatment. |
| 5 | offer, salary, job, pay, process, position, long, low, offered, get | Focuses on job offers, including salary negotiations, the overall offer process, and the positions offered to candidates. |

**Table 4 Panel B.** Classification of the BERT search based on whether the interview experience was negative, positive or neutral (code as neg exp, pos exp and neutral exp respectively).

|  | (1) BERT (Total) | (2) BERT (Neg Exp) | (3) BERT (Pos Exp) | (4) BERT (Neutral Exp) |
|---|---|---|---|---|
| Number of Ghost Jobs | 56559 | 36262 | 7896 | 12103 |
| % of Ghost Jobs | 21 | 74.8 | 4.64 | 24.8 |

**Table 5.** Number and Percentages of ghost jobs using the BERT model, sorted by how many employees the company has. The *% of ghost jobs* is calculated by taking the number of interview reviews indicative of ghost jobs divided by total number of interview reviews of companies in that employee number bracket. Variable definitions are detailed in Table 2.

|  | Number of Ghost Jobs | % of Ghost Jobs |
|---|---|---|
| BERT (1 to 50 Employees) | 314 | 14.5 |
| BERT (51 to 200 Employees) | 480 | 16.8 |
| BERT (201 to 500 Employees) | 216 | 23.9 |
| BERT (501 to 1000 Employees) | 756 | 23 |
| BERT (1001 to 5000 Employees) | 1.52e+04 | **24.8** |
| BERT (5001 to 10000 Employees) | 1.3e+04 | 22.7 |
| BERT (10000+ Employees) | 2.58e+04 | 18.7 |
| BERT (Unknown) | 857 | 19.8 |

**Table 6** Ranking of the Industries - the top 5 industries with the highest percentage of interview reviews indicative of ghost jobs and the industries with the lowest percentage of interview reviews indicative of ghost jobs, using both keyword search and BERT models. The upper panel shows the top industries and the bottom panel, the industries with the lowest percentages, and % of ghost jobs refer to the number of interview reviews indicative of ghost jobs divided by the total number of interview reviews for companies in that particular industry. Variable definitions are detailed in Table 2.

| Top Industries (Keyword Search) | Number of Ghost Jobs (Keyword Search) | % of Ghost Jobs (Keyword Search) |
|---|---|---|
| Top Industries (BERT) | Number of Ghost Jobs (BERT) | % of Ghost Jobs (BERT) |
| Publishing | 572 | 34.4 |
| Internet & Web Services | 2.2e+03 | 30.1 |
| Grantmaking & Charitable Foundations | 170 | 29.9 |
| Commercial Equipment Services | 102 | 29.8 |
| Software Development | 1.83e+03 | 29.8 |
| Bottom Industries (BERT) | Number of Ghost Jobs (BERT) | % of Ghost Jobs (BERT) |
| Restaurants & Cafes | 1.32e+03 | 13.6 |
| National Agencies | 550 | 13.2 |
| Accounting & Tax | 252 | 12.5 |
| Religious Institutions | 26 | 12.4 |
| Commercial Printing | 25 | 12.2 |

**Table 7.** Do word count and skill of the job affect whether the job ad is more likely to be a ghost job? This table shows a fixed effect OLS regression on the dependent variable of whether the review is indicative of a ghost job using the BERT model. Word Count refers to the word count of the interview review, experience can be either positive, neutral or negative, with positive being 1, neutral being 0 and negative being −1. Size refers to the 5 levels of the company sizes indicated by the company data. Unknown company sizes are discarded. Skill refers to whether the job is high-skilled or low-skilled. Standard errors, shown in parentheses, are clustered at the year level. Variable definitions are detailed in Table 2.

|  | (1) BERT | (2) BERT | (3) BERT |
|---|---|---|---|
| Word Count | 0.002*** | 0.002*** | 0.002*** |
|  | (0.000) | (0.000) | (0.000) |
| Experience | −0.302*** | −0.298*** | −0.298*** |
|  | (0.006) | (0.007) | (0.007) |
| Size | 0.019*** | 0.000 | 0.000 |
|  | (0.006) | (.) | (.) |
| Size$^2$ | −0.003*** | 0.000 | 0.000 |
|  | (0.001) | (.) | (.) |
| Skill | 0.030*** | 0.029*** | 0.029*** |
|  | (0.002) | (0.002) | (0.002) |
| Company FE | No | Yes | Yes |
| Year Fe | No | No | Yes |
| adj $r^2$ | 0.458 | 0.467 | 0.467 |
| N | 194809 | 194807 | 194807 |

*, **, *** represent significance at the 10%, 5%, and 1% level.

# Appendix

## A1. Examples of interview reviews that may indicate "ghost jobs"

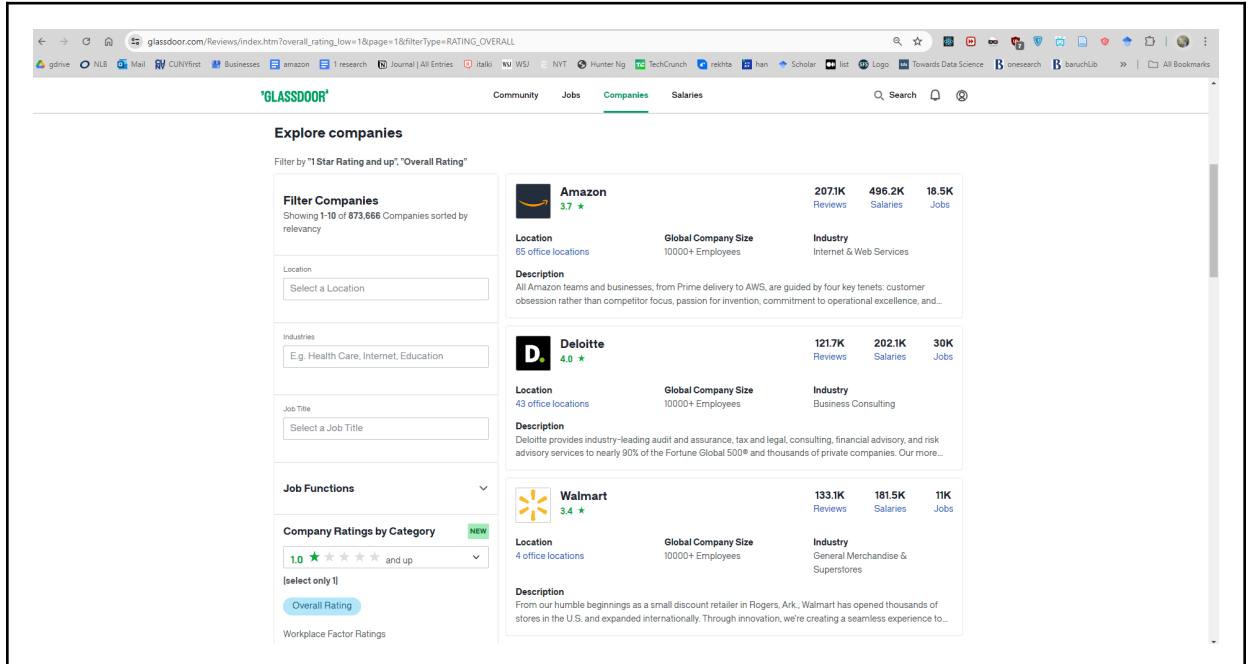

**Figure A1.** A screenshot of the *Explore Companies* section, where all companies that are listed on Glassdoor are displayed here. There are 873,666 companies that have a overall 1 start or up rating. These are sorted by relevance to the general user and randomly displayed across all industries and company size. I retrieve randomly a representative sample of 1,203 companies. *glassdoor.com*

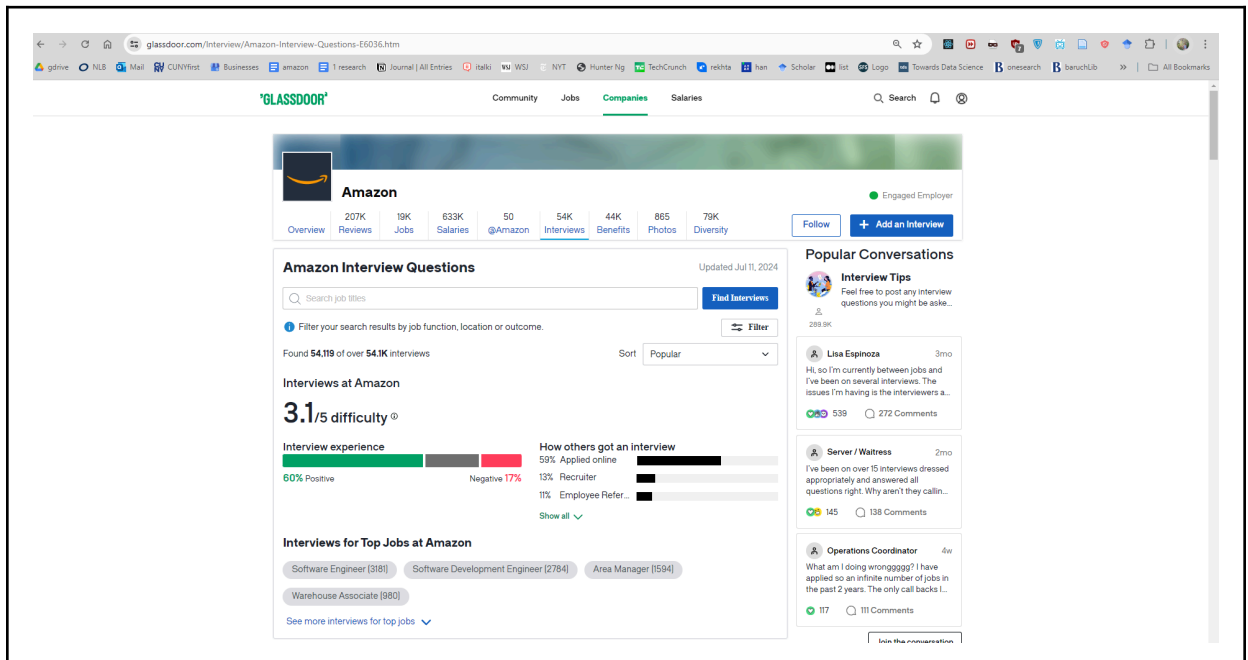

**Figure A2.** A screenshot of the *Interviews* section, where job-seekers can post their experiences about their job interviews. Glassdoor has stated that these interviews will never be taken down on request by the implicated firm, to signal their commitment to transparency on Glassdoor. *glassdoor.com*

*"The recruiter didn't call at our scheduled time, I emailed her asking if that was still a good time for her and no response. 15 minutes later she emails me saying she couldn't reach me, BS. I reschedule for later that day and it was clear why she avoided calling earlier. She was not prepared and had almost no info on the role. She proceeded to tell me that the hiring manager picked our resume out of the bunch, blah blah, all to keep me interested after I told her I didn't want to move forward with more interviews unless it was serious ( **they tried filling this role last year and put it on hold bc "they didn't know what they were looking for**"-recruiter. Also, I'd been interviewing for months after being laid off and was exhausted). I then met with the hiring manager, I had a great convo and she seemed very excited. I completed the final round and met with the team. I waited a couple days and then reached out to the recruiter to check for updates, and surprise she ghosted me. Not even the decency to respond to a candidate that completed their effing interviews, disgusting."*

**Figure A3.** An interview experience posted on 12th April 2023 for the position of Product Designer at 2U. The bolded words indicate that this could be a "ghost job" as they could be gathering a pipeline of candidates in anticipation of current employees quitting their jobs.

> *"The interview process was pretty quick because at the time I was actively interviewing with another company and was pretty far into interviews. I really appreciated that they came together to get me in quick. As part of their interview process, they DID ask me to spend almost an entire day interviewing with several different people on the potential team and hiring manager. In some ways, it is nice to get to know several people, but on the other hand, as a candidate, **it's exhausting to have interviews every half hour for 5 hours**. That to me seems unnecessary. They did also give me a surprise scenario to test whether or not I knew what I was talking about. Would have appreciated more feedback about how interviews Int. In the end, I was a bit ghosted and **they didn't even update me on having cancelled their position until I reached out to them**."*

**Figure A4.** An interview experience posted on 10th June 2019 for the position of Change Management Support at ADM. The bolded words indicate that this could be a "ghost job" and HR is getting other departments to interview candidates to possibly get a sense of the market.

## A2. Textual Analysis

I first use the *langdetect* library in Python, which is a port of Google's language-detection library and uses statistical methods to identify the language of a given text. It relies on pre-trained language models that use n-grams to make predictions about which language the text is in. Table A1 shows the breakdown.

| Table A1. Breakdown of the number of reviews initially in each language in the dataset. I use only the English reviews - "en" for the rest of the article. | | | |
|---|---|---|---|
| en | 269347 | pt | 4315 |
| fr | 3593 | es | 3148 |
| de | 1554 | it | 1451 |
| nl | 373 | unknown | 292 |
| af | 190 | ca | 36 |
| da | 13 | cy | 11 |
| ro | 8 | tl | 8 |
| so | 8 | sv | 5 |
| no | 5 | id | 5 |
| pl | 4 | sl | 3 |
| fi | 3 | sk | 2 |
| et | 2 | hr | 1 |
| tr | 1 | vi | 1 |

For the keyword search, I used the following list of words compiled through manual observation and research from the news articles on this phenomenon. I tested with a final list of 192 phrases which could represent a review that was indicative of a ghost job.

| Table A2. List of keywords used for the keyword search to check whether an interview review is indicative of a ghost job. | | |
|---|---|---|
| weren't hiring | position was put on hold | no intention of hiring |
| position had been filled | no positions were available | collect resumes |
| not actually hiring | no actual vacancy | just interviewing |
| fake job post | no real job | not actively hiring |
| position canceled | position withdrawn | never intended to hire |
| not filling the position | job ad was a sham | not recruiting |
| advertisement only | job on hold | no open positions |
| recruiting freeze | job freeze | no real vacancies |
| ghosted | always accepting applications | placeholder job |
| fake listing | fake vacancy | interview without intent to hire |
| no intent | no intention | job posting expired |
| phantom job | phony job ad | resume collector |
| job no longer available | fake hiring process | dummy position |
| not a real opening | job ad scam | no job to fill |
| job not real | position never existed | job posting hoax |
| interviewed but no job | position indefinitely on hold | fake opportunity |
| never contacted back | no follow-up after interview | interview for non-existent job |
| job ad fraud | position already filled | job ad deception |
| no intent to recruit | position no longer open | no job offer |
| ghosted by employer | interviewed for closed position | interview with no feedback |
| no vacancies at this time | job listing scam | interviewed but position not available |
| job application black hole | interviewed but no response | no job at the end |
| interviewed but never hired | recruiting for non-existent job | job offer never materialized |
| job application trap | no job interviews | position disappeared |
| resume gathering | interviewed with no intent to hire | bogus job ad |
| false job advertisement | fake recruitment process | interviewing for show |
| job no longer exists | resume solicitation | position on indefinite hold |
| job ad just for publicity | fake job vacancy | interviewing without hiring |
| interview for nonexistent position | no plans to hire | not actually recruiting |
| job interview deception | phantom hiring | position doesnt exist |
| no position available | interview with no job | no real job offer |
| interview for phantom position | no plans to recruit | job application scam |
| no recruitment | dummy vacancy | not recruiting actively |
| position permanently closed | no active hiring | position not open |
| sham job ad | no longer hiring | ad with no hiring intention |
| always hiring without positions | no active vacancies | job freeze in place |
| recruitment stopped | job ad with no positions | position not existent |

| application trap | interviewed but nothing | position already taken |
| --- | --- | --- |
| job freeze announcement | job ad without positions | no real hiring process |
| phantom job ad | position withdrawn permanently | ad without actual hiring |
| interview without positions | application but no job | no active recruitment |
| interviewed but no roles | no real job posting | position unavailable |
| resume collection ad | interviewed for non-existent role | job ad permanently closed |
| position unavailable permanently | no follow-up interview | interview without job offer |
| no follow-up contact | resume collection without jobs | position filled but still interviewing |
| no recruitment intention | no roles available | fake job advertisement |
| no vacancies advertised | ghosted job application | job ad without hiring |
| phantom vacancy | interview with no open positions | no positions open |
| dummy job advertisement | job not open | position freeze |
| no hiring intentions | position closed | interview with no result |
| no hiring after interview | interviewed without jobs | no recruitment process |
| no available positions | sham vacancy | fake recruitment |
| position fake | no follow-up hiring | resume collection for no job |
| no job positions | vacancy not real | interview but no jobs |
| interview without vacancy | position on hold indefinitely | job posting scam |
| no hiring after application | ghost application process | no recruitment after interview |
| fake job listing | phantom job process | no job hiring |
| interviewed with no result | no recruitment intention | position closed indefinitely |
| resume collection scam | fake application process | no follow-up after application |
| interviewed without vacancy | no job recruitment | application without jobs |
| fake job role | position not open for hiring | application but no hiring |
| resume collection for no vacancy | | |

For the classification of whether a interview review was indicative of a ghost job or not, I employed the keyword search in Table A2, and I also trained a BERT model to the ghost jobs. Since there may be a lot of nuances in the language that could indicate whether a review was a ghost job or not, a deep-learning model may be a better approach. I used ChatGPT-4o, the most advanced AI model currently. I used the prompt -

*"Ghost jobs refer to the phenomenon where employers put out job ads but are not actually hiring. I will give a list of interview reviews with numbers. If you are 90% sure the interview review indicates a ghost job, say yes; otherwise, say no. No explanation needed."*

on 2,000 randomly selected interview reviews, and then I trained a BERT model to . The specifications for the BERT model are listed in Table A3.

| Table A3. List of specifications for the training of the BERT model | |
|---|---|
| num_train_epochs | 3 |
| per_device_train_batch_size | 16 |
| per_device_eval_batch_size | 16 |
| warmup_steps | 500 |
| light_decay | 0.01 |
| save_steps | 1 |
| evaluation_strategy | epoch |

## A3. Informational Gain through Ghost Jobs

### 3.1 Model Setup

To set up my model, I rely on the costly information acquisition framework (Grossman and Stiglitz, 1980; Morris and Shin, 2002; Sims, 2003; Veldkamp, 2023). I first define several key variables that represent the post of ghost jobs in my empirical study.

**1. Company Size**: Let a set of companies $C$ be differentiated by size $S$. The size of a company is denoted by $S_i$, where $S_i$ can be small $S$, medium $M$, or large $L$.

**2. HR Department Size**: Let each company have a HR department size denoted by $H_i$, which scales with company size $S_i$. Assume $H_i = \alpha S_i$, where $\alpha$ is a positive constant. Thus, larger companies tend to have larger HR departments.

**3. Ghost Jobs**: Let $G_i$ represent the number of ghost jobs posted by company $i$. Companies post ghost jobs to acquire information about the job market, denoted by $I(G_i)$, a function of $G_i$.

**4. Cost of Posting Ghost Jobs**: The cost function for posting ghost jobs is denoted by $C(G_i, H_i)$. The cost increases with the number of ghost jobs and decreases with a larger HR department. Assume the cost function is given by:

$$C(G_i, H_i) = \beta G_i^2 - \gamma H_i G_i$$

where $\beta$ and $\gamma$ are positive constants. $\beta G_i^2$ represents the increasing marginal cost of handling more applications, and $\gamma H_i G_i$ represents efficiency gained from a larger HR department.

**5. Information Value**: The value of information obtained from ghost jobs is denoted by $V(I(G_i))$, which is an increasing but concave function:

$$V(I(G_i)) = \delta \ln(1 + G_i)$$

where $\delta$ is a positive constant representing the value derived from the information gathered.

## 3.2 Firm's Optimization Problem

Each company seeks to maximize its net benefit from posting ghost jobs. The net benefit function $B(G_i)$ for company $i$ is:

$$B(G_i) = V(I(G_i)) - C(G_i, H_i)$$

Substituting the functions, we get:

$$B(G_i) = \delta \ln(1 + G_i) - (\beta G_i^2 - \gamma H_i G_i)$$

To find the optimal number of ghost jobs $G_i^*$ that maximizes $B(G_i)$, I take the first-order condition by setting the derivative of $B(G_i)$ with respect to $G_i$ to zero:

$$\frac{dB}{dG_i} = \frac{\delta}{1 + G_i} - 2\beta G_i + \gamma H_i = 0$$

$$\delta = (1 + G_i)(2\beta G_i - \gamma H_i)$$

Solving,

$$G_i^* = \frac{-(\gamma H_i - 2\beta) + \sqrt{(\gamma H_i - 2\beta)^2 + (8\beta(\delta + \gamma H_i))}}{4\beta}$$

## 3.3 How does number of ghost jobs posting varies with company size?

**Small Companies** $S$: Small companies have a smaller $H_i$. Thus,

- $\gamma H_i$ is relatively low.
- $\sqrt{(\gamma H_i - 2\beta)^2 + (8\beta(\delta + \gamma H_i))}$ is dominated by $2\beta$.
- Smaller $G_i^*$ because the marginal cost of handling additional ghost jobs outweighs the benefits.

**Medium-Sized Companies** $M$: Medium-sized companies have a moderate $H_i$. Thus,

- $\gamma H_i$ is larger than for small companies but not as large as for big companies.
- The term $(\gamma H_i - 2\beta)^2$ may be close to zero, reducing the negative impact in the quadratic solution, thus increasing $G_i^*$.
- The balance between information acquisition benefits and handling costs is optimal, resulting in a higher $G_i^*$.

**Large Companies** $L$: Large companies have a large $H_i$i. Thus,

- $\gamma H_i$ is large.
- $\sqrt{(\gamma H_i - 2\beta)^2 + (8\beta(\delta + \gamma H_i))}$ is affected more by $\gamma H_i$.
- Marginal value of additional information is low, leading to a smaller increase in $G_i^*$ compared to medium-sized companies.

### 3.4 Comparative Statics

**Effect of HR Department Size** $H_i$:

- When $H_i$ is small (small companies), the term γHigamma H_iγHi is small, making $G_i^*$ low.
- When $H_i$ is moderate (medium-sized companies), $G_i^*$ is maximized because costs are balanced, and information value remains significant.
- When $H_i$ is large (large companies), the costs are lowered, but diminishing returns on information value cap the increase in $G_i^*$.

### 3.5 FOC Conclusion

By solving the first-order condition, I show that:

- Medium-sized companies post more ghost jobs than both small and large companies. This is because they have sufficient HR capacity to handle the costs associated with ghost jobs but still derive significant value from the additional information.
- Small companies post fewer ghost jobs due to limited HR resources and high marginal cost
- Large companies also post fewer ghost jobs because they face diminishing marginal returns on information due to other ways of acquiring market intelligence

**A4. Latent Dirichlet Allocation to establish rigor of Glassdoor data**

several natural language processing (NLP) techniques. Firstly, I use Latent Dirichlet Allocation (LDA) to determine the topics inherent in the interview reviews. LDA is a generative probabilistic model used for topic modeling in natural language processing (NLP). LDA assumes that each interview review is a mixture of several topics, and each topic is characterized by a distribution of words. The generative process involves sampling a topic distribution for each review, then sampling a topic for each word in the review, and finally sampling the word from the topic-specific distribution. The model parameters are then inferred using variational Bayes to reduce the Kullback-Leibler (KL) distance among topics. LDA confirms that the data from Glassdoor is high-quality.

ChatGPT-4o achieves accurate text classification through its sophisticated architecture, primarily built on the transformer model introduced by Vaswani et al. (2017). The transformer architecture consists of multiple layers of self-attention mechanisms and feed-forward neural networks, which enable the model to capture intricate relationships and dependencies within the given job interview review.